\documentclass{statsoc}
\usepackage[a4paper]{geometry}
\usepackage{amsmath}
\usepackage{graphicx,psfrag,epsf,amsfonts,amscd, amssymb,latexsym,verbatim}
\usepackage{enumerate}
\usepackage[english]{babel}
\usepackage{natbib}
\usepackage{url}

\def\bSig\mathbf{\Sigma}

\title[Model Projections in Model Space]{Multi-model inference through projections in model space}
\author[J.M. Ponciano]{Jos\'e Miguel Ponciano}
\address{Department of Biology, University of Florida, Gainesville, FL, 32611, USA.}
\email{josemi@ufl.edu}
\author[J.M. Ponciano and M.L. Taper]{Mark L. Taper}
\address{Department of Ecology, Montana State University, Bozeman, MT, 59717, USA.}

\begin{document}

\begin{abstract}
Information criteria have had a profound impact on modern ecological science. They allow researchers to estimate which probabilistic approximating models are closest to the generating process. Unfortunately, information criterion comparison does not tell how good the best model is. Nor do practitioners routinely test the reliability (e.g. error rates) of  information criterion-based model selection. In this work, we show that these two shortcomings can be resolved by extending a key observation from Hirotugu Akaike's original work. Standard information criterion analysis considers only the divergences of each model from the generating process. It is ignored that there are also estimable divergence relationships amongst all of the approximating models. We then show that using both sets of divergences, a model space can be constructed that includes an estimated location for the generating process. Thus, not only can an analyst determine which model is closest to the generating process, she/he can also determine how close to the generating process the best approximating model is. Properties of the generating process estimated from these projections are more accurate than those estimated by model averaging. The applications of our findings extend to all areas of science where model selection through information criteria is done.
\end{abstract}
\keywords{error rates in model selection, Kullback-Leibler divergence, model projections, model averaging, Akaike's Information Criterion}

\section{Introduction}
\label{sec:intro}
Recent decades have witnessed a remarkable growth of statistical ecology as a discipline, and today, stochastic models of complex ecological processes are the hallmark of the most salient publications in ecology \citep{zeng2018neutral,gravel2016stability,leibold2004metacommunity}.  Entropy and the Kullback-Liebler divergence as instruments of scientific inquiry are now at the forefront of the toolbox of quantitative ecologists, and many exciting new opportunities for their use are constantly being proposed (e.g. \citet{milne2017horton,roach2017entropy, kuricheva2017radiative, fan2017entropies,casquilho2017discussing, cushman2018calculation}). One of the most important, but under explored,  applications of the Kullback-Liebler divergence remains the study or characterization of the error rates incurred while making model selection according to information criteria \citep{taper2016evidential}. This research is particularly relevant when, as it almost always happens in science, none of the candidate models exactly corresponds to the chance mechanism generating the data.  

Understanding the impact of misspecification of statistical models constitutes a key knowledge gap in statistical ecology, and many other areas of biological research for that matter (e.g. \citet{yang2018bayesian}). Research by us and many others (see citations in \citet{taper2016evidential}) has led to detailed characterizations of how the probability of making the wrong model choice using any given information criterion, not only may depend on the amount of information (\textit{i.e.} sample size) available, but also on the degree of model misspecification.  

Consequently, in order to estimate the error rates of model selection according to any information criterion, practitioners are left with the apparent ``catch-22'' of being able to estimate how likely it is to erroneously deem as best that model which is furthest apart from the generating model, only after having accomplished the unsolved task of estimating the location of the candidate models relative to the generating process and to each other.  

In this paper, we propose a solution to this problem.  Our solution was motivated by the conceptualization of models as objects in a multi-dimensional space as well as a re-interpretation of Akaike's (1973) seminal AIC paper using elementary geometry.  Starting from Akaike's geometry, we show how to construct a model space that includes not only the set of candidate models but also an estimated location for the generating process. Now, not only can an analyst determine which model is closest to the generating process, she/he can also determine the (hyper)spatial relationships of all models and how close to the generating process the best model is.     

In 1973, Hirotugu Akaike wrote a truly seminal paper developing the AIC. Although Akaike referred to the AIC as ``An Information Criterion'', the AIC is universally known as ``Akaike Information Criterion''. Various technical accounts deriving the AIC exist in the literature (see for instance the general derivation of \citet{burnham2004multimodel}, Chapter 7), but few clarify every single step of the mathematics of Akaike's derivation (but see  \citet{deleeuw1992}). Although focusing on the measure-theoretic details, deLeeuw's account makes it clear that Akaike's paper was a paper about ideas, more than a paper about a particular technique. Years of research on this project has led us to understand that only after articulating Akaike's ideas, the direction of a natural extension of his work is easily revealed and understood. Therefore, as an introduction for our results we want to focus the first part of this paper on reviewing these key ideas of Akaike.  Although thinking of models and the generating mechanism as objects with a specific location in space is mathematically challenging, this exercise may also prove to be of use to study the adequacy of another common statistical practice in multi-model inference: model averaging.  

Intuitively, if one thinks of the candidate models as a cloud of points in a Euclidean space, then it would only make sense to ``average'' the model predictions if the best approximation of the generating chance mechanism in that space is located somewhere inside the cloud of models.  If however the generating model is located outside such cloud, then performing model average will only at best, worsen the predictions of the closest models to the generating mechanism.  The question then is, can this idea of thinking about models as points in a given space be mathematically formalized? Can the structure and location of the candidate models and the generating mechanism be somehow estimated and placed in a space? If so, then the answer to both questions above (\textit{i.e.} the error rates of multi-model selection under misspecification and when shall an analyst perform model averaging) could be readily explored.  These questions are the main motivation behind the work presented here.  

\section{The AIC and a natural geometric extension: Model Projections in Model Space} 
Technical accounts deriving Akaike's Information Criterion (AIC) exist in the literature (see for instance the general derivation of Burnham and Anderson 2002, Chapter 7), but few have attempted to clarify Akaike's 1973 paper, step by step.  A notable exception is \citet{deleeuw1992} introduction to \citet{akaike1973} Information Theory, which made it clear that more than a technical mathematical statistics paper, Akaike's seminal contribution was a paper about ideas: ``$\ldots$This is an `ideas' paper,' promoting a new approach to statistics, not a mathematics paper concerned with the detailed properties of a particular technique$\ldots$''  deLeeuw then takes on the task of extracting the ideas from the technical probabilistic details and takes on the task to come up with a unified account clarifying both, the math and the ideas involved.  His account is important because it makes evident that at the very heart of Akaike's derivation was a geometrical use of Pythagoras' theorem.   Our contribution is to take the derivation one step further by using Pythagoras' theorem again to attain not a relative, but an absolute measure of how close each model in a model set is from the generating process. In what follows we will set the stage to explain our contribution using \citet{akaike1973}, \citet{akaike1974} and \citet{deleeuw1992}. Our account will focus more on the ideas than in the technical, measure theoretic details for the sake of readability and also because this approach will allow us to shift directly to the core of our contribution.

Akaike's 1973 paper is difficult and technical but at the same time, it is a delightful reading because he managed to present his information criterion as the natural consequence of a logical narrative.  That logical narrative consisted of six key insights that we string together here to arrive at  what we believe is a second natural consequence of Akaike?s foundational thoughts.  In this paper, we present these insights followed by our extension to Akaike's work to reach a method for model projection in model space, along with a detailed worked out example.  

\subsection{Theoretical insights from Akaike (1973)}
Akaike's quest seems to have been motivated by a central objective in scientific practice: trying to come up with some measure of comparison between an approximating model and the generating model.  Following Akaike, we shall be concerned for the time being with the parametric situation where the probability densities are specified by a set of parameters $\theta =({{\theta }_{1}},{{\theta }_{2}},\ldots ,{{\theta }_{L}})'$ in the form $f(x,\theta)$. The true, generating model will be specified by setting $\theta=\theta_{0}$ in the density $f$. 
\subsubsection{\textit{Insight 1: Discrepancy from truth can be measured by the average of some function of the likelihood ratio}}
Akaike's first important insight follows from two observations.  First, under the parametric setting  defined above, the comparison between a general model and the truth can be done via the likelihood ratio, or some function of the likelihood ratio. Second, because the data $X$ are random, the average discrimination over all possible data would better represent the distance between a model and the truth.  Such an average would then be written as 
$$
\mathcal{D}(\theta ,{{\theta }_{0}};\Phi )=\int{f}(x;{{\theta }_{0}})\Phi (\tau (x,\theta ,{{\theta }_{0}}))dx={{\mathbb{E}}_{X}}\left[ \Phi (\tau (X,\theta ,{{\theta }_{0}})) \right],
$$
\noindent where the expectation is over the sampled stochastic process of interest, $X$. We denote the likelihood ratio as $\tau (x,\theta ,{{\theta }_{0}})=\frac{f(x;\theta )}{f(x;{{\theta }_{0}})}$ and a twice differentiable function of it as $\Phi(\tau (x,\theta ,{{\theta }_{0}}))$. .

Akaike then proposed to study under a general framework the sensitivity of this average discrepancy to the deviation of $\theta$ from $\theta_{0}$.    

\subsubsection{\textit{Insight 2: $\mathcal{D}(\theta ,{{\theta }_{0}};\Phi )$ is scaled by Fisher's Information Matrix}}
Akaike thought of expanding the average discrepancy $\mathcal{D}(\theta ,{{\theta }_{0}};\Phi )$ via a Taylor series around $\theta_{0}$ and keep a second order approximation.  Akaike's second insight then consisted on noting the strong link between such approximation and the theory of Maximum Likelihood (ML).

For a univariate $\theta$, the Taylor series approximation of the average function $\Phi$ of the likelihood ratio is written as

\begin{equation}\label{TSKL}
\mathcal{D}(\theta ,{{\theta }_{0}};\Phi )\approx \mathcal{D}({{\theta }_{0}},{{\theta }_{0}};\Phi )+(\theta -{{\theta }_{0}}){{\left. \frac{\partial \mathcal{D}(\theta ,{{\theta }_{0}};\Phi )}{\partial \theta } \right|}_{\theta ={{\theta }_{0}}}}+\frac{{{(\theta -{{\theta }_{0}})}^{2}}}{2!}{{\left. \frac{{{\partial }^{2}}\mathcal{D}(\theta ,{{\theta }_{0}};\Phi )}{\partial {{\theta }^{2}}} \right|}_{\theta ={{\theta }_{0}}}}+\ldots 
\end{equation}

After some calculations (see Appendix), this second order approximation is given by $\mathcal{D}(\theta ,{{\theta }_{0}})\approx \Phi (1)+\frac{1}{2}{\Phi }''(1){{(\theta -{{\theta }_{0}})}^{2}}\mathcal{I}({{\theta }_{0}})$, where $\mathcal{I}({{\theta }_{0}})$ is Fisher's information. Thus, the average discrepancy between an approximating and a generating model is scaled by the inverse of the theoretical variance of the Maximum Likelihood estimator, regardless of the form of the function $\Phi()$.  
\subsubsection{\textit{Insight 3: Setting $\Phi(t) = -2\log t$ connects $\mathcal{D}(\theta ,{{\theta }_{0}};\Phi )$ with Entropy and Information Theory}}
Akaike proceeded to arbitrarily set the function $\Phi(t)$ to  $\Phi(t) = -2\log t$.  Using this function not only furthered the connection with ML theory, but also introduced the connection of his thinking with Information Theory. By using this arbitrary function, the average discrepancy becomes a divergence because $\mathcal{D}({{\theta }_{0}},{{\theta }_{0}})=\Phi (1)=0$ and the approximation of the average discrepancy, heretofore denoted as $\mathcal{W}(\theta ,{{\theta }_{0}})$, is now directly scaled by the theoretical variance of the Maximum Likelihood estimator: $\mathcal{D}(\theta ,{{\theta }_{0}})\approx \mathcal{W}(\theta ,{{\theta }_{0}})={{(\theta -{{\theta }_{0}})}^{2}}\mathcal{I}({{\theta }_{0}})$. For a multivariate $\theta_{0}$ we get then that $\mathcal{W}(\theta ,{{\theta }_{0}})=(\theta -{{\theta }_{0}})'\mathcal{I}({{\theta }_{0}})(\theta -{{\theta }_{0}})$ where $\mathcal{I}({{\theta }_{0}})$ is Fisher's Information matrix. Conveniently then, the arbitrary factor of 2 gave his general average discrepancy function the familiar `negentropy' or Kullback-Leibler (KL) divergence form
\begin{align}
  & \mathcal{D}(\theta ,{{\theta }_{0}})=-2\int{f}(x;{{\theta }_{0}})\log \left( \frac{f(x;\theta )}{f(x;{{\theta }_{0}})} \right)dx \nonumber\\ 
 & \quad \quad \quad \ =-2{{\mathbb{E}}_{X}}\left[ \log \frac{f(X;\theta )}{f(X;{{\theta }_{0}})} \right] \nonumber\\ 
 & \quad \quad \quad \ =-2\left[ {{\mathbb{E}}_{X}}\left( \log f(X;\theta ) \right)-{{\mathbb{E}}_{X}}\left( \log f(X;{{\theta }_{0}} \right) \right] \nonumber\\ 
 & \quad \quad \quad \ =2{{\mathbb{E}}_{X}}\left( \log f(X;{{\theta }_{0}}) \right)-2{{\mathbb{E}}_{X}}\left( \log f(X;\theta ) \right) = KL(\theta,\theta_{0})
\end{align}
\noindent thus bringing together concepts in ML estimation with a wealth of results in Information Theory. The two expectations (integrals) in the last line of the above equation were often succinctly denoted by Akaike as $Sgg$ and $Sgf$ respectively: these are the neg-selfentropy and the neg-crossentropy terms.  Thus he would write that last line as $KL(\theta,\theta_{0})= 2[Sgg -Sgf]$.

\subsubsection{\textit{Insight 4: $\mathcal{D}(\theta ,{{\theta }_{0}})$ is  minimized at the ML estimate of $\theta$.}}
Aikaike's fourth critical insight was to note that a Law of Large Numbers (LLN) approximation of the KL divergence between the generating process and an approximating model is minimized when the candidate model is evaluated at its model parameters ML estimates.  Such conclusion can be arrived at even if the generating stochastic model is not known.  Indeed, given a sample of size $n$, $X_{1},X_{2},\ldots, X_{n}$ from the generating model, from the LLN we have that 
$$
\hat{\mathcal{D}}_{n}(\hat{\theta},\theta_{0}) = -2\times\frac{1}{n}\sum_{i=1}^{n}{\rm log}\frac{f(x_{i};\hat{\theta})}{f(x_{i};\theta_{0})},
$$
\noindent which is minimized at the ML estimate $\hat{\theta}$. Akaike actually thought that this observation could be used as a \textit{justification for the maximum likelihood principle}: ``Though it has been said that the maximum likelihood principle is not based on any clearly defined optimum consideration, our present observation has made it clear that it is essentially designed to keep minimum the estimated loss function which is very naturally defined as the mean information for discrimination between the estimated and the true distributions.'' \cite{akaike1973}.

\subsubsection{\textit{Insight 5: Minimizing $\mathcal{D}(\theta ,{{\theta }_{0}})$ is an average approximation problem.}}
Akaike's fifth insight was to recognize the need to account for the randomness in the ML estimator.  Because multiple realizations of a sample $X_{1},X_{2},\ldots, X_{n}$ yield multiple estimates of $\theta$, one should in fact think of the average discrepancy as a random variable, where the randomness is with respect to distribution of the MLE $\hat{\theta}$. Let $\mathcal{R}({{\theta }_{0}})={{\mathbb{E}}_{{\hat{\theta }}}}\left[ \mathcal{D}(\hat{\theta },{{\theta }_{0}}) \right]$ denote our target average (average over the distribution of $\hat{\theta}$. Then, the problem of minimization of the KL divergence then becomes a problem of approximation of the average
$$
\begin{array}{ccc}
   \mathcal{R}({{\theta }_{0}})={{\mathbb{E}}_{{\hat{\theta}}}}\mathcal{D}(\hat{\theta },{{\theta }_{0}}) & = & 2{{\mathbb{E}}_{{\hat{\theta }}}}\left[ {{\mathbb{E}}_{X}}\left( \log f(X;{{\theta }_{0}}) \right)-{{\mathbb{E}}_{X}}\left( \log f(X;\hat{\theta })|\hat{\theta } \right) \right]  \\
   & = & 2{{\mathbb{E}}_{X}}\left( \log f(X;{{\theta }_{0}}) \right)-2{{\mathbb{E}}_{{\hat{\theta }}}}\left[ {{\mathbb{E}}_{X}}\left( \log f(X;\hat{\theta })|\hat{\theta } \right) \right].  \\
\end{array}
$$
The first term in this expression is an unknown constant whereas the second term is a double expectation.

\subsubsection{\textit{Insight 6:  $\mathcal{D}(\theta,\theta_{0})$ can be approximated geometrically using Pythagoras' theorem.}}
Instead of estimating the expectations above, Akaike thought of substituting the probabilistic entropy $\mathcal{D}(\hat{\theta },{{\theta }_{0}})$ with its Taylor Series approximation $\mathcal{W}(\hat{\theta} ,{{\theta }_{0}})=(\hat{\theta} -\theta_{0})'\mathcal{I}({{\theta }_{0}})(\hat{\theta} -{{\theta }_{0}}),$ which can then be interpreted as a squared statistical distance. This is the square of a statistical distance because proximity between points is weighted by the dispersion of the points in the multivariate space, which is in turn proportional to the eigenvalues of the positive definite matrix $\mathcal{I}({{\theta }_{0}})$.  This sixth insight led him straight into the path to learning about the KL divergence between a generating process and a set of proposed probabilistic mechanisms/models.  By viewing this quadratic form as a statistical distance, Akaike was able to use a battery of  clear measure-theoretic arguments relying on various convergence proofs to derive the AIC.  

Interestingly, and although he doesn't explicitly mentions it in his paper, his entire argument can be phrased geometrically. Expressing the average discrepancy as the squared of a distance was a crucial step in Akaike's derivation because it opened the door for its decomposition using Pythagoras theorem.  By doing such decomposition, one can immediately visualize through a simple sketch the ideas in his proof. We present such sketch in Figure 1.  In that figure, the key triangle with a right angle has as vertices the truth $\theta_{0}$ of unknown dimension $L$, the ML estimator $\hat{\theta}$ of dimension $k \leq L$, denoted $\hat{\theta}_{k}$  and finally, $\theta_{0k}$ which is the orthogonal projection of the truth in the plane where all estimators of dimension $k$ lie, that we will denote $\Theta_{k}$ (Figure 1a). Figure 1b shows a fourth crucial point in this geometrical interpretation:  it is the estimator of $\theta_{0}$ from the data using a model with the same model form than the generating model, but with parameters estimated from the data.  To distinguish it from $\hat{\theta}_{k}$ we denote this estimator $\hat{\theta}_{0}$. This estimator can be thought of as being located in the same model plane as the generating model $\theta_{0}$. Akaike's LLN approximation of the KL divergence as an average of log-likelihood ratios $
\hat{\mathcal{D}}_{n}(\hat{\theta},\theta_{0}) = -2\times\frac{1}{n}\sum_{i=1}^{n}{\rm log}\frac{f(x_{i};\hat{\theta})}{f(x_{i};\theta_{0})}
$  comes to play in this geometric derivation as the edge labeled $e^{2}$ in Figure 1b that traces the link between $\hat{\theta}_{0}$ and the ML estimator $\hat{\theta}_{k}$.  Following Akaike's derivation then, the ML estimator $\hat{\theta}_{k}$ can be thought as the orthogonal projection of $\hat{\theta}_{0}$ onto the plane $\Theta_{k}$. 
  
Before continuing with our geometric interpretation, we alert the reader that in Figure 1 we have labeled all the edges with a lowercase letter with the purpose of rendering this geometric visualization as simple as possible. In what follows we will do the algebraic manipulations with these letters, and although in so doing we run the unavoidable risk of trivializing one of the greatest findings of modern statistical science, we do it all for the sake of transmitting the main idea behind his proof.  The reader however, should be well aware that these edges (lower case letters) denote for the most part random variables and that in the real derivation, more complex arguments including limits in probability and fundamental probability facts are needed. 

\begin{figure}
\begin{center}
\makebox{\includegraphics[angle=-90,width=1\linewidth]{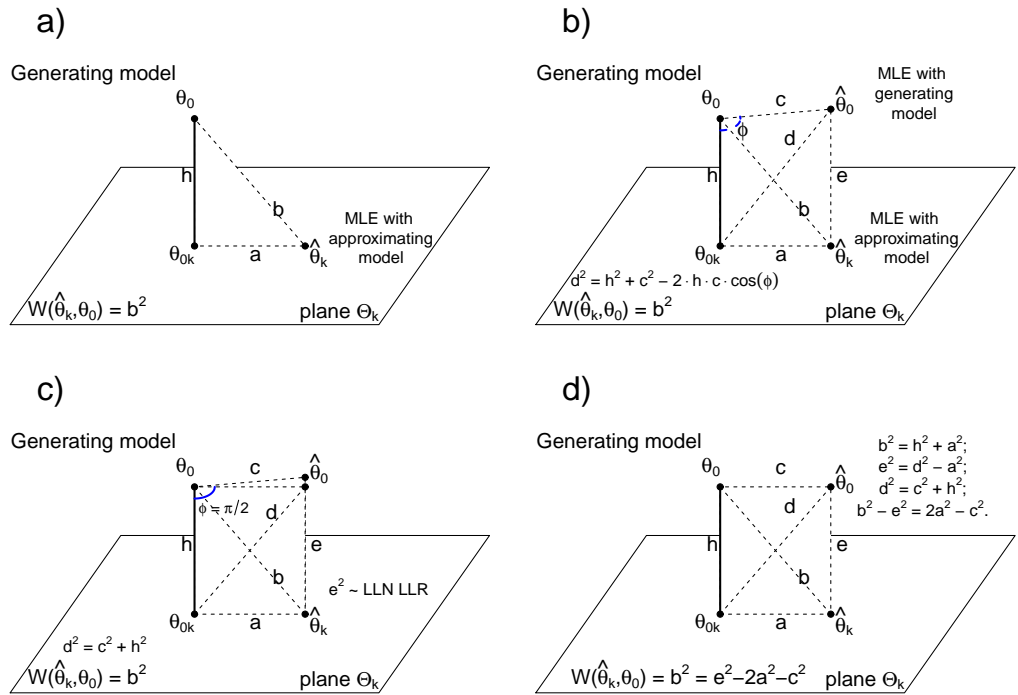}}
\end{center}
\caption{The geometry of Akaike's Information Criterion. Panel \textbf{a)} shows $\theta_{0}$, which is the generating model and $\theta_{0k}$ which is the orthogonal projection of the generating model into the space $\Theta_{k}$ of dimension $k$.  $\hat{\theta}_{k}$ is the ML estimate (MLE) of an approximating model of dimension $k$ given a data set of size $n$. Akaike's objective was to solve for $b^{2}$, which represents in this geometry $\mathcal{W}(\hat{\theta} ,{{\theta }_{0}})$, the quadratic form approximation of the divergence between the generating and the approximating models. Akaike showed that $\hat{\theta}_{k}$ can be thought of as the orthogonal projection of the MLE of $\hat{\theta}_{0}$ (panel \textbf{b)}). This last quantity $\hat{\theta}_{0}$ represents the MLE of $\theta_{0}$ with a finite sample of size $n$ and assuming that the correct model form is known. The angle $\phi$ is not necessarily a right angle, but Akaike used $\phi\approx\pi/2$ so that the generalized Pythagoras theorem (equation on the lower left side of panel \textbf{b}) could be approximated with the simple version of Pythagoras (equation on the lower left side of panel \textbf{c}) when the edge $h$ is not too long. When implemented, this Pythagoras equation can be used in conjunction with the other Pythagorean triangles in the geometry to solve for the squared edge $b$.  The equations leading to such solution are shown in panel \textbf{d} }
\end{figure}

In simple terms then, the objective of this geometric representation is to solve for the square of the edge length $b$, whose square is in fact the quadratic form approximation of the KL divergence between the generating process and the approximating model.  That is, $b^{2} = \mathcal{W}(\hat{\theta} ,{{\theta }_{0}})$. Proceeding with our geometric interpretation, note that the angle $\phi$ between edges h and c in Figure 1b is not by necessity a right angle, and that the generalized Pythagoras Theorem to find the edge length $d$ applies.  Akaike then noted that provided that the approximating model is in the vicinity of the generating mechanism, the third term of the generalized Pythagoras form of the squared distance $d^{2} = c^{2} + h^{2} - 2ch\cos{\phi}$ remained insignificant compared with the other squared terms, and so he proceeded to simply use only the first two terms, $c^{2}$ and $h^{2}$. (See Figure 1c). The staggering and successful use of the AIC in the scientific literature shows that such approximation is in many cases reliable.  This approximation allowed him to write the squared distance $d^{2}$ in two different ways: as $d^{2} \approx c^{2} + h^{2}$ and as $d^{2} = a^{2} + e^{2}$. Because by construction, we have that $b^{2} = h^{2} + a^{2}$, one can immediately write the difference $b^{2}-e^{2}$ as 
\begin{eqnarray*}
b^{2}-e^{2} & = & h^{2} + a^{2} - d^{2} + a^{2}\\
 & = & h^{2} + a^{2} - c^{2} - h^{2} + a^{2},  
\end{eqnarray*}

\noindent and then solve for $b^2$ (see Figure 1 d): 
\begin{equation}\label{Pythag.out}
b^{2} = e^{2} + 2a^{2} - c^{2}.
\end{equation}   
Using asymptotic expansions of these squared terms, the observed Fisher's information and using known convergence in probability resutls, Akaike showed when multiplied by the sample size $n$, the difference of squares $c^{2} - a^{2}$ was approximately chi-squared distributed with degrees of freedom $L-k$ and that $na^{2} \sim \chi^{2}_{k}$.  Then, multiplying equation \ref{Pythag.out} by $n$ gives
$$
n{{b}^{2}}=n\mathcal{W}({{\hat{\theta }}_{k}},{{\theta }_{0}})\approx \underbrace{n{{\mathcal{D}}_{n}}({{{\hat{\theta }}}_{k}},{{{\hat{\theta }}}_{0}})}_{=\text{log-likelihood ratio}}+\underbrace{n{{a}^{2}}}_{\sim\chi _{k}^{2}}-\underbrace{n({{c}^{2}}-{{a}^{2}})}_{\sim\chi _{L-k}^{2}}.
$$
\noindent The double expectation from the original average discrepancy definition is then implemented by simply replacing the chi-squares by their expectations, which immediately gives
$$
n{{\mathbb{E}}_{{{{\hat{\theta }}}_{k}}}}\left[ \mathcal{W}({{{\hat{\theta }}}_{k}},{{\theta }_{0}}) \right]\approx n{{\mathcal{D}}_{n}}({{\hat{\theta }}_{k}},{{\hat{\theta }}_{0}})+2k-L,\,\text{or}
$$
\begin{equation}\label{crude.aic}
{{\mathbb{E}}_{{{{\hat{\theta }}}_{k}}}}\left[ \mathcal{W}({{{\hat{\theta }}}_{k}},{{\theta }_{0}}) \right]\approx \frac{-2}{n}\sum\limits_{i=1}^{n}{\log f}({{x}_{i}};{{\hat{\theta }}_{k}})+\frac{2k}{n}-\frac{L}{n}+\frac{2}{n}\sum\limits_{i=1}^{n}{\log f}({{x}_{i}};{{\hat{\theta }}_{0}}).
\end{equation}
\noindent Note that the final form of the AIC only includes the terms $-2\sum\limits_{i=1}^{n}{\log f}({{x}_{i}};{{\hat{\theta }}_{k}})+2k$ in that expression.  To understand why only these two terms can be used to achieve multi-model comparison as usually conceived (see Figure 2),  recall first that what equation \ref{crude.aic} is in fact approximating is 

\begin{figure}
\begin{center}
\makebox{\includegraphics[angle=-90,width=0.95\linewidth]{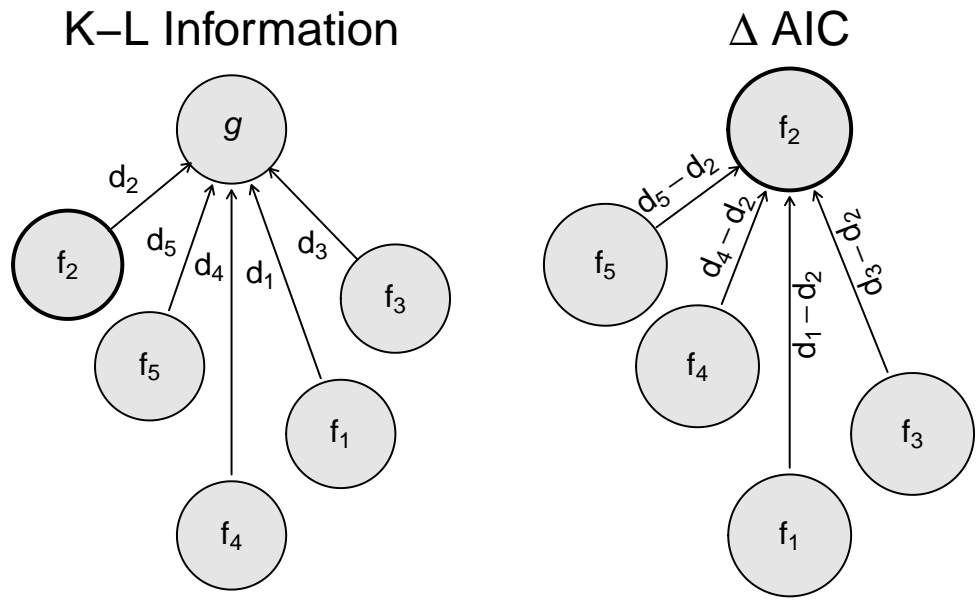}}
\end{center}
\caption{Schematic representation of the logic of multi-model selection using the AIC. $g$ represents the generating model and $f_{i}$ the $i^{th}$ approximating model.  The Kullback-Leibler information discrepancies ($d_i$) are shown on the left as the distance between approximating models and the generating model.  The $\Delta$AICs shown on the right measures the distance from approximating models to the best approximating model.  All distances are on the information scale.}
\end{figure}

\begin{equation}
\mathcal{R}({{\theta }_{0}})={{\mathbb{E}}_{{\hat{\theta }}}}\mathcal{D}({{\hat{\theta }}_{k}},{{\theta }_{0}})=-2{{\mathbb{E}}_{{\hat{\theta }}}}\left[ {{\mathbb{E}}_{X}}\left( \log f(X;{{{\hat{\theta }}}_{k}})|{{{\hat{\theta }}}_{k}} \right) \right]+2{{\mathbb{E}}_{X}}\left( \log f(X;{{\theta }_{0}}) \right)
\end{equation}
\noindent which is the expected value with respect to $\hat{\theta}_{k}$ of 
\begin{equation}
-2\int{f}(x;{{\theta }_{0}})\log \frac{f(x;{{{\hat{\theta }}}_{k}})}{f(x;{{\theta }_{0}})}dx=-2\int{f}(x;{{\theta }_{0}})\log f(x;{{\hat{\theta }}_{k}})dx+2\int{f}(x;{{\theta }_{0}})\log f(x;\theta_{0} )dx.
\end{equation}

Akaike concluded that an unbiased estimation of the expected value over the distribution of $\hat{\theta}_{k}$ of the first integral above (Akaike's $Sgf$) would be given by the average of the first two terms in equation $\hat{\theta}_{k}$. The first term in equation \ref{crude.aic} is $-2/n$ times the log likelihood with the approximating model. The last two terms cannot be known, but because upon comparing various models they will remain the same can be ignored in practice. Because n also remains the same across models, in order to compare an array of models one only has to compute $AIC = -2\sum\limits_{i=1}^{n}{\log f}({{x}_{i}};{{\hat{\theta }}_{k}})+2k$ and choose the model with the lower score as the one with the smallest discrepancy to the generating model.  The logic stemming from Akaike's reasoning and used to date can be graphically represented by Figure 2 (redrawn from \citet{burnham2011aic}). One of our central motivations to write this paper is the following:  by essentially ignoring the remainder terms in equation \ref{crude.aic}, since  1973 practicioners have been almost invariably selecting the ``least worst'' model among a set of models.  In other words, we as a scientific community, have largely disregarded the question of how far, \textit{in absolute terms not relative}, is the generating process from the best approximating model.  Suppose the generating model is in fact very far from all the models in a set of models currently being examined. Then, the last two terms in equation \ref{crude.aic} will be very large with respect to the first two terms for all the models in a model set that is being examined, and essentially any differences between the terms $-2\sum\limits_{i=1}^{n}{\log f}({{x}_{i}};{{\hat{\theta }}_{k}})+2k$  for every model will be meaningless.  

Finally, we wish to point out that in the ``popular'' statistical literature within the Wildlife Ecology sciences, (\textit{e.g.} \citet{burnham2004multimodel,burnham2011aic}) it is often asserted that  $-AIC/2$ is an estimator of ${{\mathbb{E}}_{{\hat{\theta }}}}\left[ {{\mathbb{E}}_{X}}\left( \log f(X;{{{\hat{\theta }}}_{k}})|{{{\hat{\theta }}}_{k}} \right) \right]$.  It is not. As \citet{akaike1974} states, the estimator of this quantity is $-AIC/2n$.  For the qualitative comparison of models, this distinction makes no difference, but factoring the sample size $n$ into the AIC allows a comparer of models to assess not only which model appears best, but what is the strength of evidence for that statement.  

\subsection{The problem of multiple models}
A model-centric view of science coupled with a disavowal of the absolute truth of any model pushes the scientist to the use of many models.  Once this stance is taken, the question of how to use multiple models in inference naturally arises.  Inference by the best model is not adequate as many models may be indistinguishable on the basis of pairwise comparison. 

Currently, the dominant method for incorporating information from multiple models is model averaging.  This comes in several flavors. In all case model averaging is inherently, and generally explicitly, a Bayesian approach.  Most common in ecology is averaging model parameter estimates or model predictions using Akaike weights.  The Akaike weight for the $i^{th}$ model is given as:

$$
{{w}_{i}}=\frac{\exp \left( {-{{\Delta }_{i}}}/{2}\; \right)}{\sum\limits_{r=1}^{R}{\exp \left( {-{{\Delta }_{r}}}/{2}\; \right)}},
$$
\noindent where $\Delta_i$  is the difference between a model's AIC value and the lowest AIC value from the model set of $R$ models indexed by $r$.  Although it is not always pointed out, the $w_i$ are posterior probabilites based on subjective priors of the form \citep{burnham2011aic}:
\begin{equation}\label{AICweights}
{{q}_{i}}=C\cdot \exp \left( \frac{1}{2}{{k}_{i}}\log \left( n \right)-{{k}_{i}} \right)
\end{equation}
\noindent where $q_i$is the prior for the $i^{th}$ model, $C$ is a normalization constant, $k_{i}$ is the number of parameters for model $i$ and $n$ is the number of observations.The use of this prior makes model averaging a confirmation approach \citep{bandyopadhyay2016belief}.

Two difficulties with model averaging for an evidentialist are: 1) the weights are based on beliefs, and are thus counter to an evidential approach. And 2) as a practical matter, model averaging does not take into account model redundancy.  The more effort put into building models in a region of model space, the more heavily that region gets weighted in the average. We propose the alternative of estimating the properties of the best projection of truth, or a generating model, to the hyper-plane containing the model set. This mathematical development extends Akaike's insight by using the known KL distances among models as a scaffolding to aid in the estimation of the location of the generating model.

For convenience, we follow Akaike's 1974 notation briefly mentioned above where $g$ denotes the generating model and $f$ the approximating model.  Then the so called cross-entropy and neg-entropy are written as
$$
Sgf=\int{f}(x;{{\theta }_{0}})\log f(x;{{\hat{\theta }}_{k}})dx \quad\text{and}\quad Sgg=\int{f}(x;{{\theta }_{0}})\log f(x;\theta )dx\quad \text{respectively.}
$$
\noindent  Akaike observed that the cross-entropy could be estimated with 

\begin{equation}\label{sgf.hat}
\widehat{Sgf}=\frac{1}{n}\sum\limits_{i=1}^{n}{\log f}({{x}_{i}};{{\hat{\theta }}_{k}})-\frac{k}{n}=-\frac{AIC}{2n}.
\end{equation}

\noindent Under the same considerations as Akaike's geometrical derivation, we now extend these ideas to the case where we want to draw inferences from the spatial configuration of $f_{1},f_{2}\ldots$ approximating models to the generating model $g$.

\subsection{A geometrical extension of Akaike's extension to the principle of Maximum Likelihood}

The fundamental idea of our contribution is to use the architecture of model space to try to estimate the projection of truth onto a (hyper)plane where all the approximating models lie.  Having estimated the location of truth, even without formulating an explicit model for it would anchor the AIC statistics in a measure of overall goodness of fit, as well as provide invaluable insights into the appropriateness of model averaging.    The intuition of the feasibility of such task comes from the realization that approximating models have similarities and dissimilarities.  A modeler is drawn naturally to speak of the space of models.  All that remains to is to realize that that language is not metaphor, but fact.  KL divergences can be calculated between any distributions and are not restricted to between generating processes and approximating models.  A set of models has an internal geometrical relationship which constrains and therefore has information about the relationship of approximating models and the generating process. 

Computational advances have rendered straightforward algorithmic steps that while conceptually feasible would have been computationally intractable at the time that Akaike was developing the AIC.  First, it is now easy to calculate the KL divergence between any two models.  For instance, for the Normal distribution, the KL discrepancy can be computed exactly using the package gaussDiff in the statistical software R.  Other packages will estimate the KL divergences of arbitrary distributions. Thus for a large set of approximating models, a matrix of estimated KL divergences among the set of models can be constructed.  Second, parallel processing has tamed the computer intensive Non-Metric Multidimensional (NMDS) scaling algorithm which can take an estimated matrix of KL divergences and estimate the best Euclidean representation of model space in a (hyper)plane with coordinates $\left( {{y}_{1}},{{y}_{2}},... \right)$. Nothing in our development restricts model space to be restricted to two-dimensions.  To emphasize this we speak of a (hyper)plane, but to have any hope of visualizing we stay in $\mathcal{R}^2$ for this paper. 

Suppose then that one can place the approximating models $f_{1},f_{2}\ldots$ on a Euclidean plane, as in the sketch below.  For simplicity we have placed only two models in the sketch.  Our derivation is not constrained to their particular configuration in the plane, relative to the generating model (truth), as the sketches a) and b) below show. Define $m$ with coordinates $(y_{1}^{\star},y_{2}^{\star})$ as the orthogonal projection of the generating model (truth) into the Euclidean plane of models. This projection is separated by the length $h$ to the generating model $g$. Define  $d(f_{i},m)$ as the distance in the hyper(plane) of model $i$ from model $m$. Of course, the edges and nodes in this plane are random variables, associated with a sampling error.  But, for the sake of simplicity and just as we did above to explain Akaike's derivation of the AIC, we conceive them for the time being as simple fixed nodes and edges. 

Then, using Pythagoras and thinking of the KL divergences as squared distances, the following equations have to hold simultaneously

$$
\left\{ \begin{matrix}
   KL(g,{{f}_{1}}) & = & d{{({{f}_{1}},m)}^{2}}+{{h}_{1}}^{2}  \\
   KL(g,{{f}_{2}}) & = & d{{({{f}_{2}},m)}^{2}}+{{h}_{2}}^{2}  \\
   {} & \vdots  & {}  \\
\end{matrix} \right.
$$

\noindent where necessarily ${{h}_{1}}={{h}_{2}}={{h}_{i}}=\ldots =h$ . In practice, one can decompose the KL divergence into an estimable component, $Sg{{f}_{i}}$ and a fixed unknown component $Sgg$.  Given that the $Sg{{f}_{i}}$ are estimable as in equation \ref{sgf.hat}, one can re-write the above system of equations including the unknown constants $Sgg,y_{1}^{\star },y_{2}^{\star }$ as follows:

\begin{equation}\label{MPeqs}
	\left\{
 \begin{array}{ccc}
   Sgg-\widehat{Sg{{f}_{1}}}-d{{({{f}_{1}},m(y_{1}^{\star },y_{2}^{\star }))}^{2}} & = & h^{2},  \\
   Sgg-\widehat{Sg{{f}_{2}}}-d{{({{f}_{2}},m(y_{1}^{\star },y_{2}^{\star }))}^{2}} & = & h^{2},  \\
 \vdots& \vdots & \vdots  \\ 
\end{array}\right. \\ 
 \end{equation}

Then, operationally, in order to estimate the location of the orthogonal projection of the generating model in the plane of approximating models, one can easily program the system of equation \ref{MPeqs} into an objective function that, for a given set of values of the unknown parameters $Sgg,y_{1}^{\star },y_{2}^{\star }$  computes the left hand sides of equation \ref{MPeqs} and returns the sum of the squared differences between all the $h_{i}^{2}$. Since $h^{2} = h_{i}^2$ for all $i$ (See Figure 3), a simple minimization of this sum of squared differences should lead to an optimization of the unknown quantities (See Figure 4). Note however that in these equations the terms $Sgg$ and $h^{2}$ appear always as a difference, and hence are not separable.  Fortunately, a nonparametric, multivariate estimate of $Sgg$ can be readily computed.  We use the estimator proposed by \citet{berrett2016efficient}, a multivariate extension of the well-known univariate estimator by \citet{kozachenko1987sample}. Other nonparametric entropy estimators could be used if they prove to be more appropriate..

\begin{figure}
\begin{center}
\makebox{\includegraphics[angle=-90,width=1.1\linewidth]{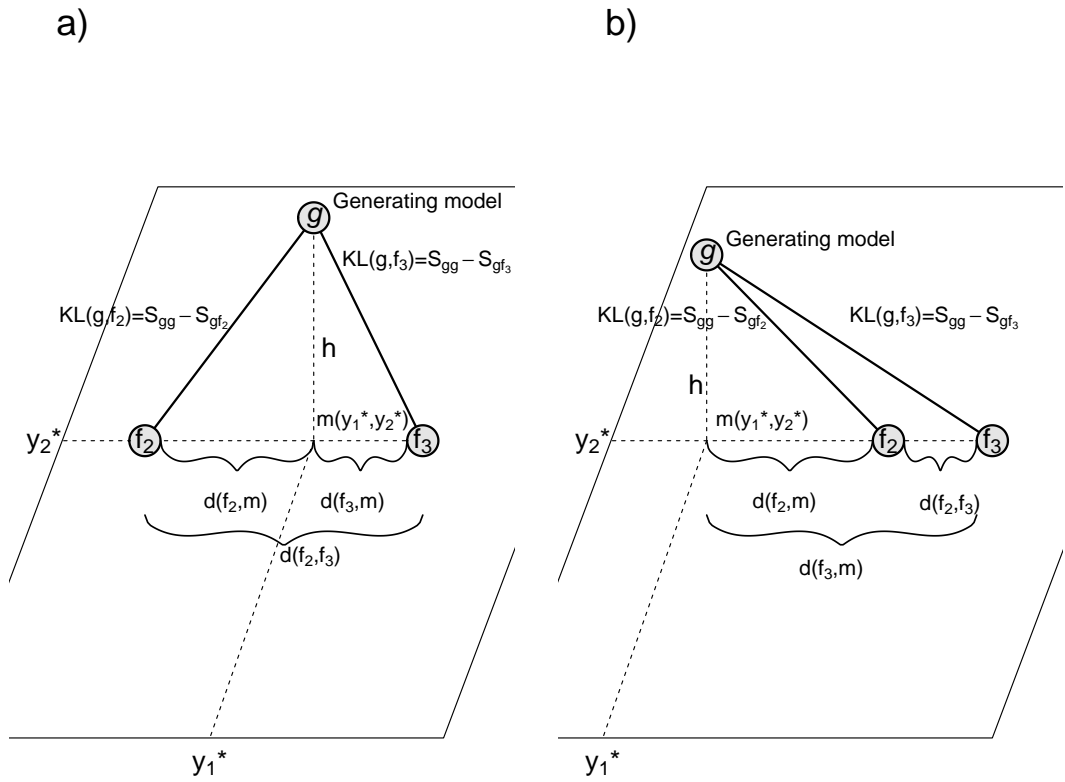}}
\end{center}
\caption{The geometry of model space.  In this figure, $f_2$ and $f_3$ are approximating models residing in a (hyper)plane. g is the generating model. m is the projection of g onto the (hyper)plane. $d(\dot,\dot)$ are distances between models in the plane. $d(f_{2},f_{3}) \approx KL(f_{2},f_{3})$ with deviations due to the dimension reduction in NMDS and non-Euclidian behavior of KL divergences. As KL divergences decrease, they become increasingly Euclidian. Panel a shows a projection when m is within the convex hull of the approximating models, and Panel b shows a projection when m is outside of the convex hull.}
\end{figure}

\begin{figure}
\begin{center}
\makebox{\includegraphics[angle=0,width=0.95\linewidth]{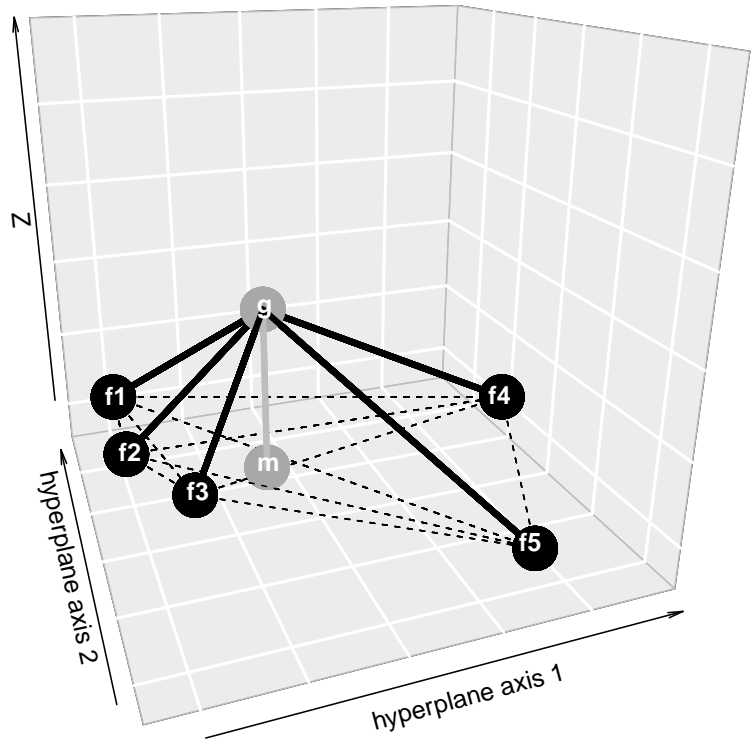}}
\end{center}
\caption{The models of Figure 2 visualized by our new methodology. As before, $g$ is the generating model and $f_{1},\ldots,f_{5}$, are the approximating models. The dashed lines are KL distances between approximating models, which can be calculated. The solid black lines are the KL distances from approximating models to the generating model, which now can be estimated. The model labeled m is the projection of the generating model to the plane of the approximating models. The solid gray line shows h, the discrepancy between the generating model and its best approximation in the NMDS plane.}
\end{figure}

\section{Examples}
\label{sec:example}
\subsection{An ecological application}
We demonstrate this approach with a simulation based on the published ecological work of \citet{grace2006structural}.  Analyzing community composition data at 90 sites over 5 years, they study the generation of plant community diversity after wild fire using structural equation models. Structural equation modeling is a powerful suite of methods facilitating the incorporation of causal hypotheses and general theoretical constructs directly into a formal statistical analysis \citep{grace2006interface, grace2008representing, grace2008structural, grace2010specification}. The final model that Grace and Keely arrive at is shown in Figure 5. The figure should be read to mean that species richness is directly influenced by heterogeneity, local abiotic conditions, and plant cover.  Heterogeneity and local abiotic conditions are themselves both directly influenced by landscape position, while plant cover is influenced by fire severity, which is influenced by stand age, which is itself influenced by landscape position.  Numbers on the arrows are path coefficients and represent the strength of influence. Our purpose in presenting this model is not to critique it or the model identification process by which it was found, but to use it as a reasonably realistic biological scenario from which to simulate.  In short, we play god using this as a known true generating process.  We consider in this analysis 41 models of varying complexity fitted to the simulated data.  They cover a spectrum from underfitted to overfitted.  

\begin{figure}
\begin{center}
\makebox{\includegraphics[angle=0,width=0.95\linewidth]{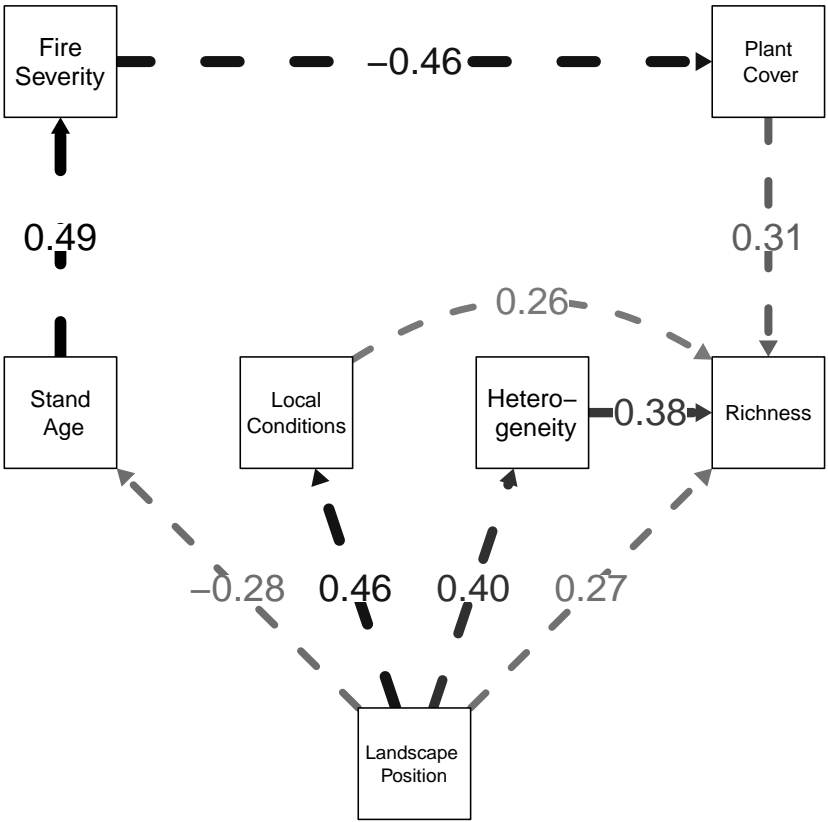}}
\end{center}
\caption{The final, simplified model explaining plant diversity from Grace and Keely (2006). Arrows indicate causal influences.  The standardized coefficients are indicated by path labels and widths.  See texts for details}
\end{figure}

We calculated the NMDS 2-dimensional model space as described above.  The stress for this NMDS is extremely low (0.006\%) indicating the model space fits almost perfectly into an $\mathcal{R}^{2}$ plane. We have plotted the fitted models in this space, grey-scale coded by the AIC categories. We have also plotted in Figure 6 the location of our methods estimated projection of the generating model to the NMDS plane, the model averaged location using Akaike weights, and the true projection of the generating model to the NMDS plane (we know this because we are acting as god). We can see in Figure 6 that the estimated projection is slightly closer to the true projection than is the model averaged location.  

\begin{figure}
\begin{center}
\makebox{\includegraphics[angle=-90,width=0.95\linewidth]{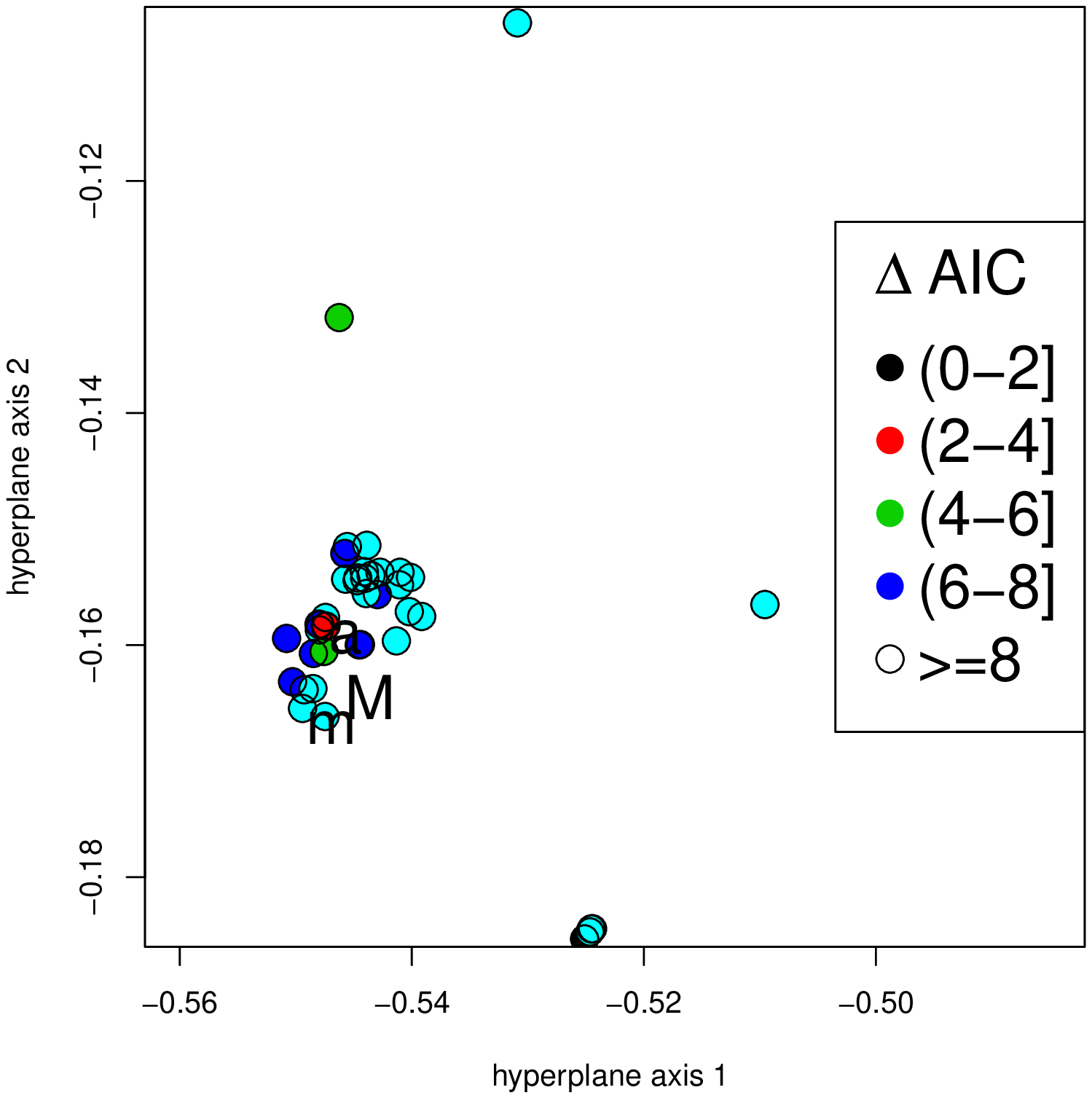}}
\end{center}
\caption{NMDS space of 41 near approximating modes. The true projection, $M$, of the generating model to the NMDS plane. The estimated location of the projection, $m$, and the location, $a$, of the model average.}
\end{figure}

In Figure 7 we plot the effect on the estimated projection and model average of deleting models from consideration.  We sequentially delete the left-most model remaining in the set recalculating locations with each deletion.  We see that the model-averaged location shifts systematically rightward with deletion, and that the location of the estimated projection is in this example more stable than the model averaged location. It remains in the vicinity of its original estimate even after all models in the vicinity of that location have been removed from consideration.  If we delete from the right, the model average moves systematically leftward.  The model projection location is, in this sequence, less stable than under deletion from the left.  These deletion exercises highlight several interesting fact about the two types of location estimates that are implicit in the mathematics, but easily over looked.  First, the model average is constrained to lie within the convex hull of the approximating model set.  If you shift the model set, you will shift the average. Second, the estimated generating model projection as a projection can lie outside of the convex hull.  Third, because of the geometrical nature of the projection estimate, models distant can contribute information to the location of the best projection. This is the difference between rightward and leftward deletion.  There are several models with high influence on the right hand side of the plot which are retained till the end in rightward deletion, but removed early in leftward deletion.

\begin{figure}
\begin{center}
\makebox{\includegraphics[angle=-90,width=0.95\linewidth]{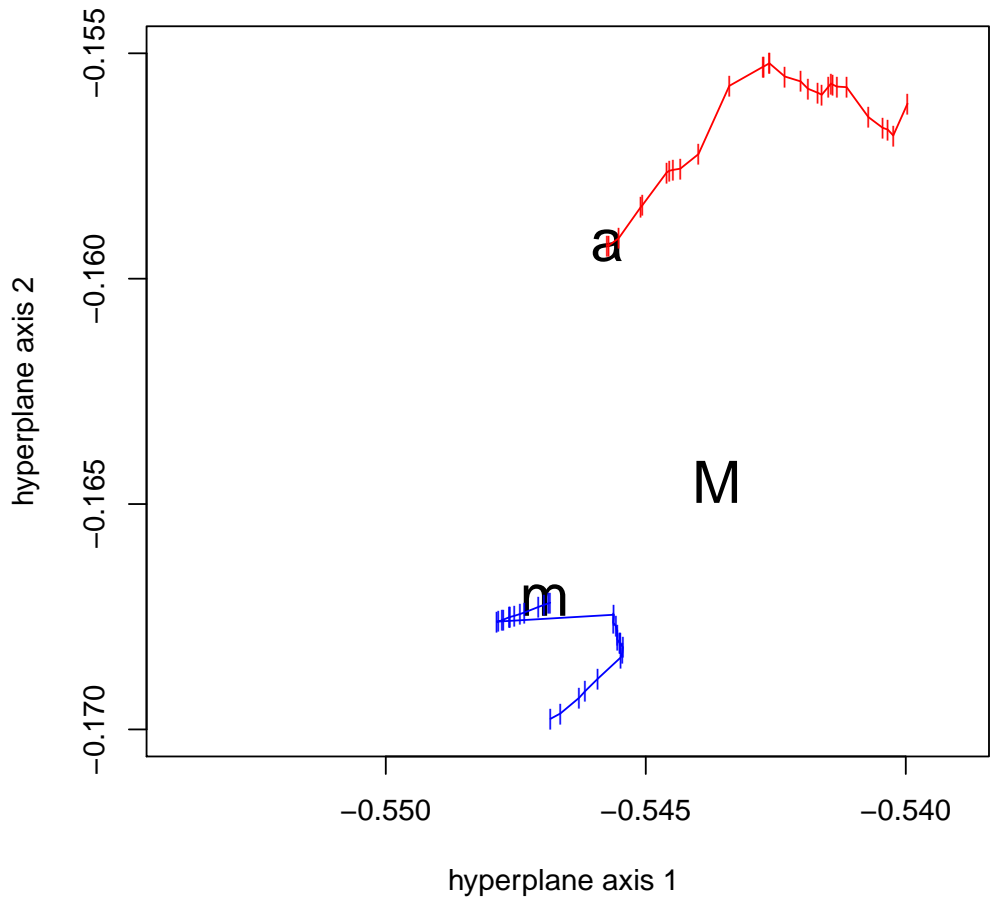}}
\end{center}
\caption{Stability test of the displacement (trajectories) of the model prediction (in blue) and the model average (in red) under deletion of $1-30$ models. $M$ denotes the true location of the orthogonal projection of the generating model in the the hyperplane. $m$ and $a$ mark the location of the model projection and the model average respectively, when the 30 models are used.  In both cases, as models are removed one by one from the candidate model set, the location of both $m$ and $a$ changes (little vertical lines). Note how the model projection estimate is more stable to changes in the model set than the model average.}
\end{figure}

Unlike model averaging, the model projection methodology also produces estimates of two more quantities.  The $Sgg$, the neg-selfentropy of the generating process is estimated as $-9.881$.  As God, we know that the true value is $-9.877$.  These two agree to three significant figures.  Also estimated is the distance of the generating process from the (hyper)plane of the NMDS model space.  This is very important, because if the generating process is far from the (hyper)plane then any property estimate based on information from the model set should be suspect.  The estimate for this discrepancy is $0.00018$, indicating that is very close the (hyper)plane. The true discrepancy is $5.8 e-08$.

\subsection{Testing the non-parametric estimation of $Sgg$}
To exemplify the independent estimation of $Sgg$ with a data set we simulated samples from a seven-dimensional multivariate normal distribution and compared the true value of $Sgg$ with its non-parametric estimate according to \citet{berrett2016efficient}.  We chose to simulate data from a multivariate normal distribution because its $Sgg$ value is known analytically. When the dimension of a multivariate normal distirbution is $p$ and is variance-covariance matrix is $\Sigma$, then
$$
Sgg = -\frac{1}{2} {\rm ln}\left\{(2\pi{\rm e})^{p} {\rm det}(\Sigma) \right\}.
$$
\noindent To carry our test, we chose 5 testing sample sizes $10,25,50,75,150$ and for each sample size we simulated 2000 data sets according to a multivariate normal distribution with $p=7$ and $\Sigma = I$, and computed each time Berrett et al.'s non-parametric estimate.  The resulting estimates, divided by the true value of $9.93257$ are plotted as boxplots in Figure 8.

\begin{figure}
\begin{center}
\makebox{\includegraphics[angle=0,width=0.85\linewidth]{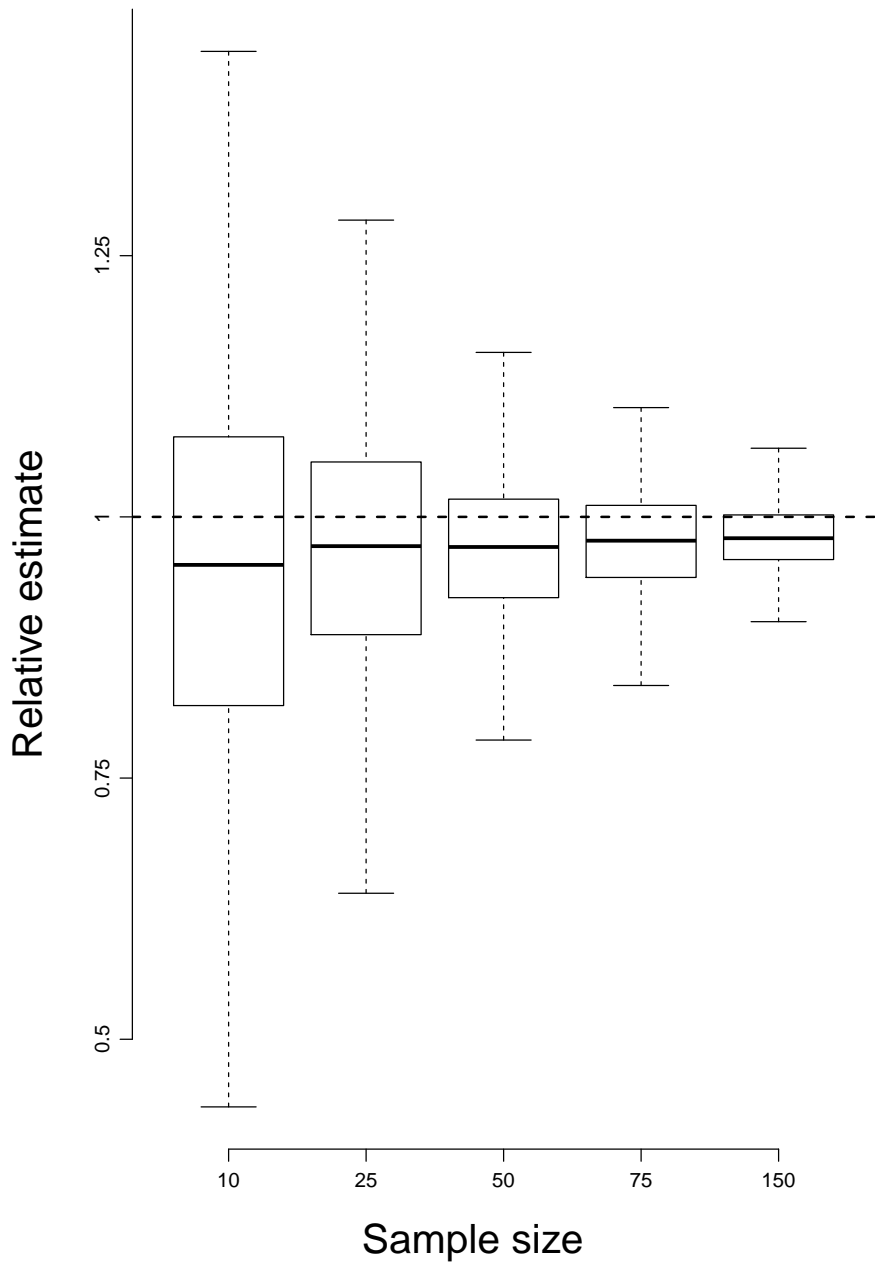}}
\end{center}
\caption{Boxplots of sets of 2000 nonparametric estimates of $Sgg$ (from Berrett et al. (2016)) relative to the true $Sgg$ value of $9.93257$,  for different sample sizes. The simulated data comes from a seven-dimensional Multivariate Normal distribution with means equal to 10 and the identity matrix as a variance-covariance matrix. The dashed, horizontal line at $1$ shows the zero-bias mark.}
\end{figure}

\section{Discussion}
\label{sec:disc}
We have constructed a novel approach to multi-model inference. Standard multi-model selection analyses only estimate the relative, not overall divergences of each model from the generating process. Typically, divergence relationships amongst all of the approximating models are also estimable (dashed lines in Figure 4). We have shown that using both sets of divergences, a model space can be constructed that includes an estimated location for the generating process (the point $g$ in Figure 4).  The construction of such model space stems directly from a geometrical interpretation of Akaike's original work. 

The approach laid out here has clear and substantial advantages over standard model identification and Bayesian based model averaging. Now the overall architecture of model space vis-a-vis the generating process is statistically estimable. Such architecture is composed of a critical set of quantities and relationships. Among these objects, we now include the estimated location of the closest projection of the generating process into the (hyper)plane of approximating models (the point $m$ in Figure 4). Second, the estimated magnitude of discrepancy between the generating process and the best projection to the model space can give the analyst an indication of whether important model attributes have been overlooked. Heuristically, the simple ability of being able to visualize model space will aid in the development of new models. 

In the information criterion literature and all scientific application,  the neg-selfentropy $Sgg$ of the generating process is simply treated as an unknown quantity. In fact, it can be estimated quite precisely as our example shows. $Sgg$ is itself of great interest because with it the overall discrepancy to the generating process becomes estimable.  Because this quantity is estimable, now the analyst can tell if any of the models in her model set are any good to begin with.  Thus, our approach solves a difficulty that has long been recognized \citep{spanos2010akaike} but yet treated as an open problem . 

With an estimated model space, the strain between a priori and post hoc inference is greatly weakened.  The study of the structure of model space gives the information to correct for misleading evidence (the probability of observing data that fails to support the best model), accommodation (over-fitting), and cooking your models (Dennis, Ponciano and Taper, in prep).  Theoretically, the more models you consider the more robust the scaffolding from which to project the location of the generating process.  Nonidentifiability and weak estimability \citep{ponciano2012assessing} are, of course, still a problem, but at least the model space approach will clearly indicate the difficulties.

Model projection is an evidential alternative \citep{taper2016evidential} to Bayesian model averaging for incorporating information from many models in a single analysis.  Model projection, because it more fully utilizes the information in the structure of model space, is able to estimate several very important quantities that are not estimated by model averaging. As we showed in our results, model projection is not as sensitive as model average to the composition of the set of candidate models being investigated. The model average does weight models with low AIC more heavily than models with higher AIC, but does not take into consideration the rate of change of model properties across the space.  Also, the estimated generating model projection is less constrained by the configuration of the model set than is the model average. 

As well as proposing solutions to existing problems, any new method also raises a variety of technical problems that need to be solved.  This is certainly the case with the model projection approach presented here. 

Currently, the model set for the model projection approach is limited by the near model requirement common to all information criteria analyses. Our exposition makes it clear that near model requirement is due to the imperfect yet useful approximation employed by Akaike while setting $\phi\approx \pi/2$ (see Figure 1).  It was only thanks to this approximation that Akaike was able to solve for the estimable divergence contrasts between all approximating models and the generating process.  This approximation breaks down in curved model spaces as the divergence from the generating process increases. Indeed, as the KL distance between approximating models and the generating model increases, $-AIC/2n$ becomes an increasingly biased and variable estimate of the $Sgf$ component of the KL distance between the approximating model and the generating model.  This effect is strong enough that sometimes very bad models can have low delta AIC values, even sometimes appearing as the best model.  The TIC and the EIC2 (\cite{konishi2008information, kitagawa2010bias}) are model identification criteria designed to be robust to model misspecification. Substituting one of these information criteria for the AIC in constructing the matrix of inter-model divergences should allow the use of models more distant from truth than is acceptable using the AIC. 

For even more distant models, it seems reasonable to think that using heteroskedastic nonlinear regressions such as described by \citet{carroll1988transformation} and \cite{carroll2006measurement} will allow for incorporating information from more distant models into the estimated projection.  If this approach does not prove effective at great distance from the generating process, at least the region in which the projection of the generating model resides can be found by plotting the density of AIC good models in the NMDS space.  The projection methodology can be applied in the high density region.

Our methodology focuses on estimation of the model space geometry but uncertainties around such estimation are not fully worked out as of yet.  As described above, one of the expectations taken in calculating the AIC is over parameter estimates. Estimation of the location and properties of the estimated projection can likely be improved using the reduced variance bias corrected bootstrap information criterion of \citet{kitagawa2010bias}.  A benefit of this is that confidence intervals on the estimated projection can be simultaneously calculated.  These intervals are based in sample space probabilities and can be expressed either as traditional confidence intervals or as evidential support intervals (see \citet{taper2016evidential}).  This contrasts with intervals produced by model averaging, which despite their sometime presentation as error statistics are actually posterior probability intervals (under the cryptic assumptions that model prior probabilities are given as in equation \ref{AICweights} and that the posterior distribution is normal). 

We think that this model projection methodology should be the starting point to do a careful, science-based inquiry of what are the model attributes that make a model a good model.  Knowing the location of the projected best model is an essential component of our multi-model development strategy because a response surface analysis can reveal what model attributes tend to be included near the location of the projected best model thus aiding in the construction of a model closer to the best projection.

\bibliographystyle{chicago}
\bibliography{PBGSSbib}

\begin{thebibliography}{}

\bibitem[\protect\citeauthoryear{Akaike}{Akaike}{1973}]{akaike1973}
Akaike, H. (1973).
\newblock Information theory as an extension of the maximum likelihood
  principle.
\newblock In B.~Petrov and F.~Csaki (Eds.), {\em Second international symposium
  on information theory}, pp.\  267--281. Budapest: Akademiai Kiado.

\bibitem[\protect\citeauthoryear{Akaike}{Akaike}{1974}]{akaike1974}
Akaike, H. (1974).
\newblock A new look at statistical-model identification.
\newblock {\em IEEE Transactions on Automatic Control\/}~{\em 19}, 716--723.

\bibitem[\protect\citeauthoryear{Bandyopadhyay, Brittan, and
  Taper}{Bandyopadhyay et~al.}{2016}]{bandyopadhyay2016belief}
Bandyopadhyay, P.~S., G.~G. Brittan, and M.~L. Taper (2016).
\newblock {\em Belief, evidence, and uncertainty: problems of Epistemic
  inference}.
\newblock Springer.

\bibitem[\protect\citeauthoryear{Berrett, Samworth, and Yuan}{Berrett
  et~al.}{2016}]{berrett2016efficient}
Berrett, T.~B., R.~J. Samworth, and M.~Yuan (2016).
\newblock Efficient multivariate entropy estimation via $ k $-nearest neighbour
  distances.
\newblock {\em arXiv preprint arXiv:1606.00304\/}.

\bibitem[\protect\citeauthoryear{Burnham and Anderson}{Burnham and
  Anderson}{2004}]{burnham2004multimodel}
Burnham, K.~P. and D.~R. Anderson (2004).
\newblock Multimodel inference: understanding aic and bic in model selection.
\newblock {\em Sociological methods \& research\/}~{\em 33\/}(2), 261--304.

\bibitem[\protect\citeauthoryear{Burnham, Anderson, and Huyvaert}{Burnham
  et~al.}{2011}]{burnham2011aic}
Burnham, K.~P., D.~R. Anderson, and K.~P. Huyvaert (2011).
\newblock Aic model selection and multimodel inference in behavioral ecology:
  some background, observations, and comparisons.
\newblock {\em Behavioral Ecology and Sociobiology\/}~{\em 65\/}(1), 23--35.

\bibitem[\protect\citeauthoryear{Carroll and Ruppert}{Carroll and
  Ruppert}{1988}]{carroll1988transformation}
Carroll, R.~J. and D.~Ruppert (1988).
\newblock {\em Transformation and weighting in regression}, Volume~30.
\newblock CRC Press.

\bibitem[\protect\citeauthoryear{Carroll, Ruppert, Stefanski, and
  Crainiceanu}{Carroll et~al.}{2006}]{carroll2006measurement}
Carroll, R.~J., D.~Ruppert, L.~A. Stefanski, and C.~M. Crainiceanu (2006).
\newblock {\em Measurement error in nonlinear models: a modern perspective}.
\newblock CRC press.

\bibitem[\protect\citeauthoryear{Casquilho and Rego}{Casquilho and
  Rego}{2017}]{casquilho2017discussing}
Casquilho, J.~P. and F.~C. Rego (2017).
\newblock Discussing landscape compositional scenarios generated with
  maximization of non-expected utility decision models based on weighted
  entropies.
\newblock {\em Entropy\/}~{\em 19\/}(2), 66.

\bibitem[\protect\citeauthoryear{Cushman}{Cushman}{2018}]{cushman2018calculation}
Cushman, S.~A. (2018).
\newblock Calculation of configurational entropy in complex landscapes.
\newblock {\em Entropy\/}~{\em 20\/}(4), 298.

\bibitem[\protect\citeauthoryear{deLeeuw}{deLeeuw}{1992}]{deleeuw1992}
deLeeuw, J. (1992).
\newblock Introduction to akaike (1973) information theory and an extension of
  the maximum likelihood principle.
\newblock In S.~Kotz and N.~L. Johnson (Eds.), {\em Breakthroughs in
  statistics}, pp.\  599--609. London: Springer.

\bibitem[\protect\citeauthoryear{Fan, Yu, He, Yu, Bai, Yang, and Wu}{Fan
  et~al.}{2017}]{fan2017entropies}
Fan, Y., G.~Yu, Z.~He, H.~Yu, R.~Bai, L.~Yang, and D.~Wu (2017).
\newblock Entropies of the chinese land use/cover change from 1990 to 2010 at a
  county level.
\newblock {\em Entropy\/}~{\em 19\/}(2), 51.

\bibitem[\protect\citeauthoryear{Grace}{Grace}{2008}]{grace2008structural}
Grace, J.~B. (2008).
\newblock Structural equation modeling for observational studies.
\newblock {\em Journal of Wildlife Management\/}~{\em 72\/}(1), 14--22.

\bibitem[\protect\citeauthoryear{Grace, Anderson, Olff, and Scheiner}{Grace
  et~al.}{2010}]{grace2010specification}
Grace, J.~B., T.~M. Anderson, H.~Olff, and S.~M. Scheiner (2010).
\newblock On the specification of structural equation models for ecological
  systems.
\newblock {\em Ecological Monographs\/}~{\em 80\/}(1), 67--87.

\bibitem[\protect\citeauthoryear{Grace and Bollen}{Grace and
  Bollen}{2006}]{grace2006interface}
Grace, J.~B. and K.~A. Bollen (2006).
\newblock {\em The interface between theory and data in structural equation
  models}.
\newblock US Geological Survey.

\bibitem[\protect\citeauthoryear{Grace and Bollen}{Grace and
  Bollen}{2008}]{grace2008representing}
Grace, J.~B. and K.~A. Bollen (2008).
\newblock Representing general theoretical concepts in structural equation
  models: the role of composite variables.
\newblock {\em Environmental and Ecological Statistics\/}~{\em 15\/}(2),
  191--213.

\bibitem[\protect\citeauthoryear{Grace and Keeley}{Grace and
  Keeley}{2006}]{grace2006structural}
Grace, J.~B. and J.~E. Keeley (2006).
\newblock A structural equation model analysis of postfire plant diversity in
  california shrublands.
\newblock {\em Ecological Applications\/}~{\em 16\/}(2), 503--514.

\bibitem[\protect\citeauthoryear{Gravel, Massol, and Leibold}{Gravel
  et~al.}{2016}]{gravel2016stability}
Gravel, D., F.~Massol, and M.~A. Leibold (2016).
\newblock Stability and complexity in model meta-ecosystems.
\newblock {\em Nature communications\/}~{\em 7}, 12457.

\bibitem[\protect\citeauthoryear{Kitagawa and Konishi}{Kitagawa and
  Konishi}{2010}]{kitagawa2010bias}
Kitagawa, G. and S.~Konishi (2010).
\newblock Bias and variance reduction techniques for bootstrap information
  criteria.
\newblock {\em Annals of the Institute of Statistical Mathematics\/}~{\em
  62\/}(1), 209.

\bibitem[\protect\citeauthoryear{Konishi and Kitagawa}{Konishi and
  Kitagawa}{2008}]{konishi2008information}
Konishi, S. and G.~Kitagawa (2008).
\newblock {\em Information criteria and statistical modeling}.
\newblock Springer Science \& Business Media.

\bibitem[\protect\citeauthoryear{Kozachenko and Leonenko}{Kozachenko and
  Leonenko}{1987}]{kozachenko1987sample}
Kozachenko, L. and N.~N. Leonenko (1987).
\newblock Sample estimate of the entropy of a random vector.
\newblock {\em Problemy Peredachi Informatsii\/}~{\em 23\/}(2), 9--16.

\bibitem[\protect\citeauthoryear{Kuricheva, Mamkin, Sandlersky, Puzachenko,
  Varlagin, and Kurbatova}{Kuricheva et~al.}{2017}]{kuricheva2017radiative}
Kuricheva, O., V.~Mamkin, R.~Sandlersky, J.~Puzachenko, A.~Varlagin, and
  J.~Kurbatova (2017).
\newblock Radiative entropy production along the paludification gradient in the
  southern taiga.
\newblock {\em Entropy\/}~{\em 19\/}(1), 43.

\bibitem[\protect\citeauthoryear{Leibold, Holyoak, Mouquet, Amarasekare, Chase,
  Hoopes, Holt, Shurin, Law, Tilman, et~al.}{Leibold
  et~al.}{2004}]{leibold2004metacommunity}
Leibold, M.~A., M.~Holyoak, N.~Mouquet, P.~Amarasekare, J.~M. Chase, M.~F.
  Hoopes, R.~D. Holt, J.~B. Shurin, R.~Law, D.~Tilman, et~al. (2004).
\newblock The metacommunity concept: a framework for multi-scale community
  ecology.
\newblock {\em Ecology letters\/}~{\em 7\/}(7), 601--613.

\bibitem[\protect\citeauthoryear{Milne and Gupta}{Milne and
  Gupta}{2017}]{milne2017horton}
Milne, B.~T. and V.~K. Gupta (2017).
\newblock Horton ratios link self-similarity with maximum entropy of
  eco-geomorphological properties in stream networks.
\newblock {\em Entropy\/}~{\em 19\/}(6), 249.

\bibitem[\protect\citeauthoryear{Ponciano, Burleigh, Braun, and Taper}{Ponciano
  et~al.}{2012}]{ponciano2012assessing}
Ponciano, J.~M., J.~G. Burleigh, E.~L. Braun, and M.~L. Taper (2012).
\newblock Assessing parameter identifiability in phylogenetic models using data
  cloning.
\newblock {\em Systematic biology\/}~{\em 61\/}(6), 955--972.

\bibitem[\protect\citeauthoryear{Roach, Nulton, Sibani, Rohwer, and
  Salamon}{Roach et~al.}{2017}]{roach2017entropy}
Roach, T.~N., J.~Nulton, P.~Sibani, F.~Rohwer, and P.~Salamon (2017).
\newblock Entropy in the tangled nature model of evolution.
\newblock {\em Entropy\/}~{\em 19\/}(5), 192.

\bibitem[\protect\citeauthoryear{Spanos}{Spanos}{2010}]{spanos2010akaike}
Spanos, A. (2010).
\newblock Akaike-type criteria and the reliability of inference: Model
  selection versus statistical model specification.
\newblock {\em Journal of Econometrics\/}~{\em 158\/}(2), 204--220.

\bibitem[\protect\citeauthoryear{Taper and Ponciano}{Taper and
  Ponciano}{2016}]{taper2016evidential}
Taper, M.~L. and J.~M. Ponciano (2016).
\newblock Evidential statistics as a statistical modern synthesis to support
  21st century science.
\newblock {\em Population ecology\/}~{\em 58\/}(1), 9--29.

\bibitem[\protect\citeauthoryear{Yang and Zhu}{Yang and
  Zhu}{2018}]{yang2018bayesian}
Yang, Z. and T.~Zhu (2018).
\newblock Bayesian selection of misspecified models is overconfident and may
  cause spurious posterior probabilities for phylogenetic trees.
\newblock {\em Proceedings of the National Academy of Sciences\/}, 201712673.

\bibitem[\protect\citeauthoryear{Zeng and Rodrigo}{Zeng and
  Rodrigo}{2018}]{zeng2018neutral}
Zeng, Q. and A.~Rodrigo (2018).
\newblock Neutral models of short-term microbiome dynamics with host
  subpopulation structure and migration limitation.
\newblock {\em Microbiome\/}~{\em 6\/}(1), 80.

\end{thebibliography}

\end{document}